# XANES spectroscopy study of Pb(Ti, Zr)O$_3$ ferroelectric thin films

Mira Mandeljc and Marija Kosec
*Jozef Stefan Institute. Jamova 39, SI-1000 Ljubljana, Slovenia*

S.P. Gabuda*, S.G. Kozlova, S. B. Erenburg and N.V. Bausk
*Institute of Inorganic Chemistry of Siberian Department of Russian Academy of Sciences Novosobirsk 630090, Russian Federation*

**Abstract.** Thin films of zirconium-rich perovskites Pb$_{1-3x/2}$La$_x$Zr$_{0.65}$Ti$_{0.35}$O$_3$ (PLZT; x=0.095) were studied using X-ray absorption near-edge structure (XANES) spectroscopy. Is shown, that slow crystallization of the system is accompanied by partial reduction of lead oxides to a metal (Pb$^0$) state. The same results were found in the samples of the same composition with addition of 30% mol of PbO. The evidence for stereo active 6s$^2$ lone-pair is discussed.

**Introduction**

Previously was studied the low-temperature crystallization of (Pb$_{0.86}$La$_{0.095}$)(Zr$_{0.65}$Ti$_{0.35}$)O$_3$ (PLZT) of ~235 - 250 nm thin films deposited on substrates made by chemical solution deposition (CSD) on platinized silicon substrates [1,2]. It was showed that, by increasing the amount of PbO excess in the film, the perowskite phase grows almost to the surface of the film [1,2]. At as low as 400 ºC, the perovskite PLZT 9.5/65/35 grows on the PbTiO3 seeding layer. The parabolic dependence of the perovskite growth at 400 ºC indicates a diffusion-controlled process.

The growth is slow: after 65 h the PLZT thin film consists of about 35 % of perovskite and 65 % of amorphous phase. The amorphous phase have a deficiency in Pb with regard to the perovskite. We used 10 mol % PbO excess. We conclude that the sublimation of PbO from the surface of the film is faster than the growth of the perovskite phase. When the limiting stoichiometry of the perovskite phase with regard to the PbO is reached, its growth is stopped. At 450 ºC, the crystallization of the perovskite phase is faster, therefore the amount of the perovskite phase after 65 h is increased in comparison to 400 ºC. However, when the limiting stoichiometry of the perovskite phase with regard to PbO is reached, the growth stops. By increasing the amount of PbO excess in the film, from the original 10 % to 30 %, the perovskite phase grows almost to the surface of the film. These results support the hypothesis that the large enough amount of the PbO is the critical factor for the crystallization of the perovskite phase even at as low as 400 or 450 °C.

From transmission electron microscopy (TEM) the perovskite and amorphous phase were determined in the samples, and amorphous phase was found to be a lead deficient. From our preliminary studies, in the samples lead is in two valence state: $Pb^{2+}$ and probably metal Pb, or even $Pb^{4+}$ or metal Pb. In this work, we studied the valence state of lead in the lead zirconate-titanate thin films by analyzing their X-ray absorption near-edge structure (XANES). The measurement results were compared with the data of density functional theory (DFT) [3] calculations.

**Experimental**

The XANES experiments were conducted at theCenter of Synchrotron Radiation Storage Ring VEPP-3 (Budker Institute of Nuclear Physics, Siberian Division, Russian Academy of Sciences) at the EXAFS station. In measurements, the storage ring operated at an energy of 2.00 GeV and a current of 50–100 mA. An ionization chamber filled with Ar/He was used as a monitoring detector. A mono block slit silicon single crystal ({111} plane) was used as a double crystal monochromator. The transmission spectra were recorded for the thin-film samples.

The $L(3)$-edge fine structure of the X-ray absorption due to the electronic dipole transitions from the $2p_{1/2}$ core level to the unoccupied upper levels was studied. The X-ray absorption edge for metallic Pb is caused by the allowed transitions to the close $6d_{5/2}$ and $7s_{1/2}$ levels. The tabulated absorption edge energies is $Pb^0 L(3) = 13.034$ keV [4]. With the aim of assigning the fine-structure components to particular electronic transitions, the transition energies $\Delta E$ were calculated for the $Pb^{2+}$ and $Pb^{4+}$ ions using DFT method [3]. In the bivalent and tetravalent lead compounds, 8 and 19 eV to higher energies shift the main peak, respectively. In addition, a fine structure appears at lower energies. The calculated fine structure of the lead in different states is confronted with experimental XANES - spectra (curves *a* in Figures 1- 6). Fine structure of the spectra is more clearly seen in the second derivatives of the X-ray absorption $L(3)$-edge curves for the lead compounds (see curves *b* in Figures 1-6).

**Results**

Generally, the excitation of the internal electrons in the X-ray absorption spectroscopy is accompanied by formation of short-living electronic hole. The typical time-living of the excited state is shorter then $10^{-15}$ s, and hence, the natural width of the spectral absorption lines is of order 5 to 10 eV. This fact explains why the resolution of XANES spectra is rather pure. Nevertheless, the numerical analysis of the spectra can help to establish the presence of definite valence states of heavy elements like Pb.

*Amorphous samples.* Were studied 2 amorphous samples: stoichiometric PLZT 9.5/65/35 thin film (Fig. 1), and of the same composition, but with 30 mol % PbO excess of PbO (Fig. 2). In spite of this difference, two spectra do not show any visible difference both in absorption and in second derivative curves. This fact correlates well with the data of previous XANES-spectroscopy study of two PbO modifications [5]. According to the data, in of both PbO and PLZT lead is divalent, and hence, their XANES spectra must be similar. This is a justification - why the excess of PbO in the last sample is not manifested in the XANES spectrum, represented in Figure 2.

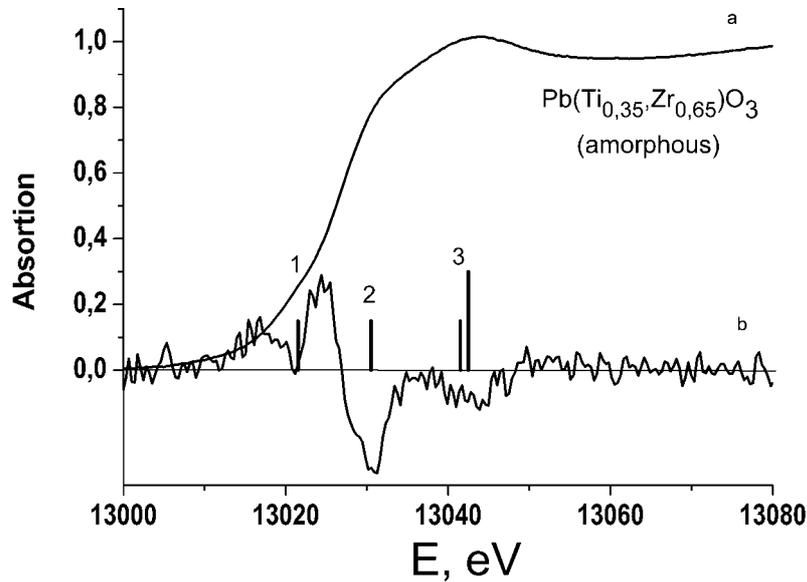

Fig. 1. The XANES Pb L(3) spectrum of stoichiometric PLZT thin film. *a* –absorption curve; *b*- its second derivative. Numbers indicates the DFT-calculated components of fine structure, caused by transitions from the basic $2p_{3/2}$ state to the $Pb^{2+}$ excited states: 1 – $6s_{1/2}$; 2- $6p_{1/2}$; 3- $7s_{1/2}$ (left) and $6d_{3/2}$.

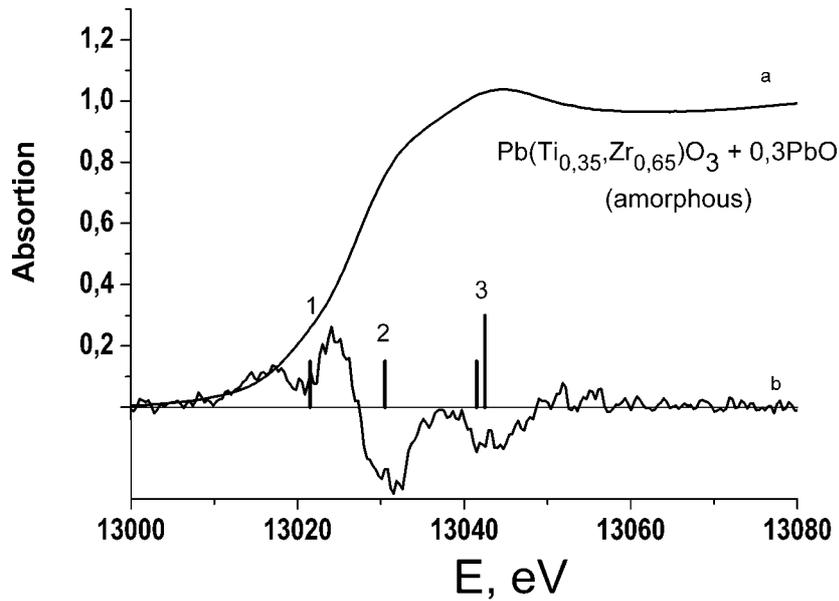

Fig. 2. The XANES Pb L(3) spectrum of PLZT thin film with 30 mol % PbO excess. The indication are the same as in Figure 1.

*Crystalline samples.* Were studied 3 crystalline samples: stoichiometric PLZT 9.5/65/35 thin film (Fig. 3), and of the same composition, but with 30 mol % PbO excess of PbO (Fig. 4). As indicated, the crystallization was conducted at 450 °C for 65 hours. Additionally, was studied the thin film sample PZT 30/70, crystallized at 600 °C for 15 minutes (Figure 5). In all three samples, lead must be on A sites in the perovskite phase - $Pb^{2+}$. The general feature of all crystalline samples is increased peak intensity in the vicinity of the energy 13 034 eV which is characteristic for a metal state of lead. From qualitative analysis of the spectra 3-5 with above discussed spectra 1-2 one can conclude that nearly 20% of lead oxides is reduced to a metal ($Pb^0$) state in all crystalline samples.

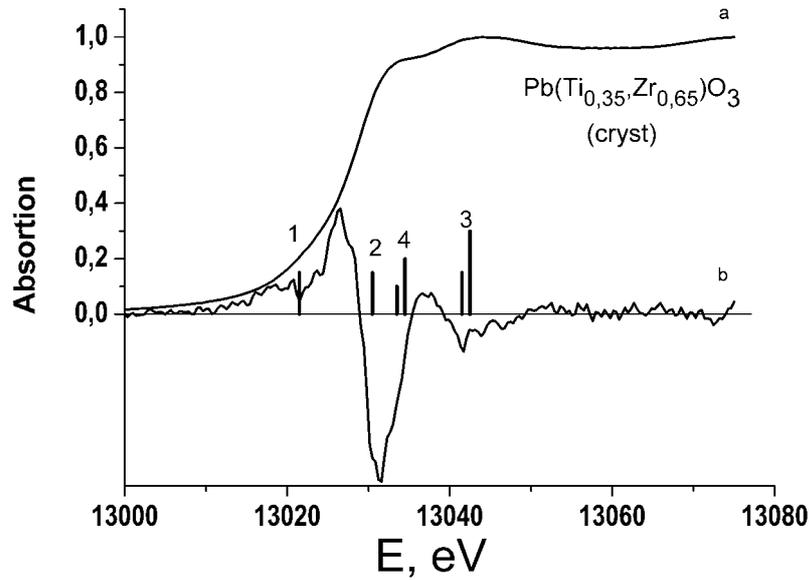

Fig. 3. The XANES Pb L(3) spectrum of crystalline stoichiometric PLZT thin film. *a* – absorption curve; *b*- its second derivative. Numbers indicates the DFT-calculated components of fine structure, caused by transitions from the basic $2p_{3/2}$ state to the $Pb^{2+}$ excited states: 1 – $6s_{1/2}$; 2- $6p_{1/2}$; 3- $7s_{1/2}$ (left) and $6d_{3/2}$; 4 – the transitions from the basic $2p_{3/2}$ state to the $Pb^0$ excited states $7s_{1/2}$ (left) and $6d_{3/2}$.

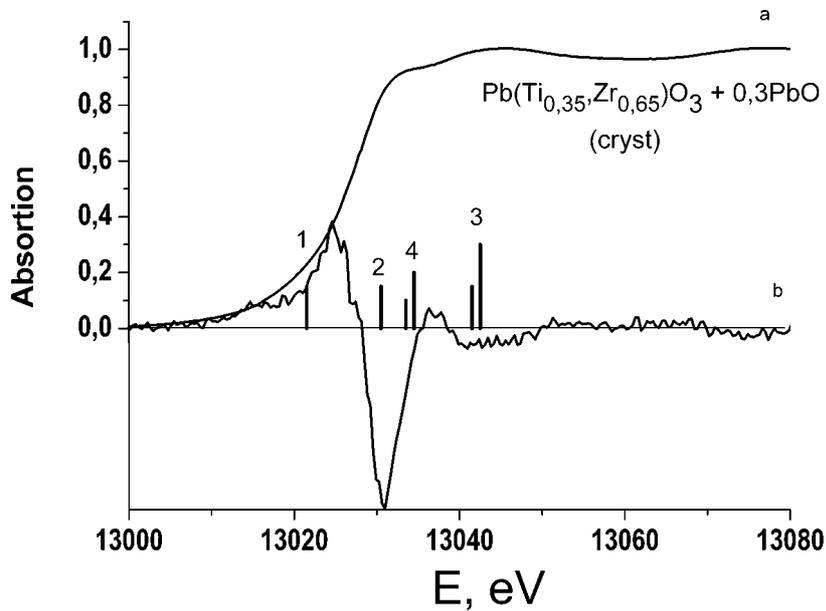

Fig. 4. The XANES Pb L(3) spectrum of crystalline PLZT thin film with 30 mol % PbO excess. The indication are the same as in Figure 3.

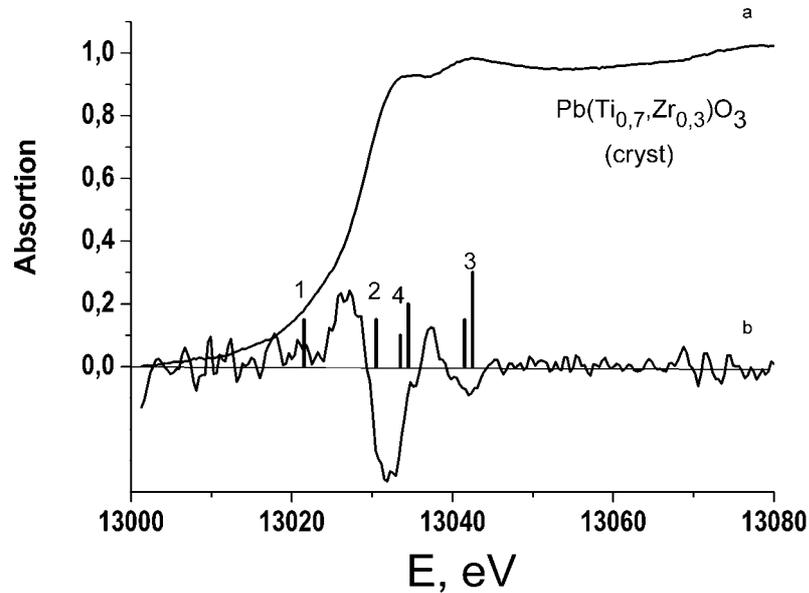

Fig. 5. The XANES Pb L(3) spectrum of crystalline PZT 30/70 thin film, crystallized at 600 °C for 15 minutes. The all indication are the same as in Figure 3.

At the same time, no one spectrum does not show any feature at the energy of 13 052 eV which is characteristic for $Pb^{4+}$. This fact definitely indicates the absence of tetravalent states of lead in all studied samples in full agreement with phase diagram of lead oxides indicating that it is impossible for lead to be $Pb^{4+}$ in used conditions.

*Activation of $6s^2$ electron pairs.* The Zr-rich samples show clear absorption in the vicinity of energy 13 021 eV (Fig. 3,4) close to the energy of $2p_{3/2} \rightarrow 6s_{1/2}$ transitions in the $Pb^{4+}$. However, this absorption should not be observed in the free $Pb^{2+}$ ions with configuration $5d^{10}6s^2$, where 6s state is filled. Previously was admitted [5] that appearance of anomalous X-ray absorption and $2p_{3/2} \rightarrow 6s_{1/2}$ transitions in the ions is evidence for a substantial change in the electronic structure of these ions in oxides, e.g., due to chemical bonding. However, the detailed band calculations did not reveal such bonding [5-7]. An alternative mechanism for the modification of electronic structure of the $Pb^{2+}$ ions can be associated with a nonzero probability of $6^{s2}$-electron tunneling through a potential barrier and with the splitting of the 6s-state.

At the same time, the titanium-rich PZT 30/70 sample do not show similar absorption in the vicinity of 13 021 eV. This fact correlates with the $^{207}Pb$ NMR solid-state data of $PbTiO_3$ [9,10] indicating inactive $6s^2$ electron pair, and the $^{207}Pb$ NMR solid-state spectra of PbZrO$_3$ (Fig. 6), showing the superposition of two lines which may be connected with activation of 6s$^2$ electron pair in this case.

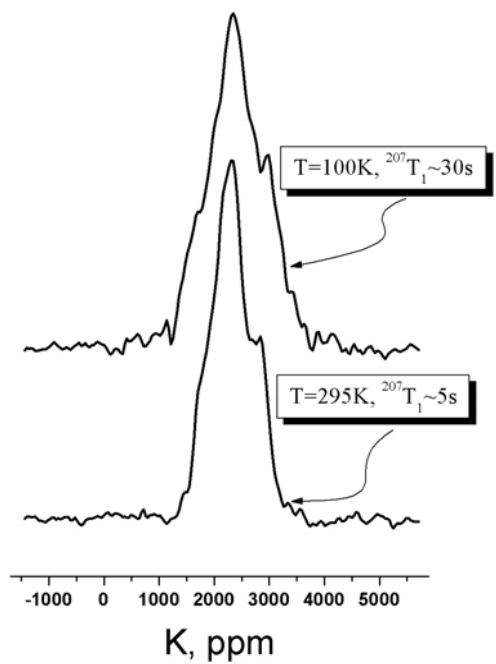

Fig.6. $^{207}$Pb NMR solid-state spectra of PbZrO$_3$ powder. Reference: Pb(CH$_3$)$_4$.

The found features of Pb(6s$^2$) electrons in lead titanate-circonate perovskites are connected probably with basic difference of PbTiO$_3$ (ferroelectric) and PbZrO$_3$ (antiferroelectric) derived from peculiarities of their clearly different electronic structures.

This work was supported by Division of chemistry and material sciences of RAS (Project No 15 of Program 4.1), and by RFBR (Grants 02-03-32816 and 02-03-32319). Support from the Ministry of Education, Science and Sports of the Republic of Slovenia within the National Research Program is also acknowledged.

*) E-mail: gabuda@casper.che.nsk.su